\documentclass[acmsmall,screen]{acmart}
\setcopyright{none}
\renewcommand{\footnotetextcopyrightpermission}[1]{}
\acmYear{2027}
\acmDOI{}
\acmISBN{}
\AddToHook{fancyhdr/before}{%
  \fancyfoot[RO,LE]{}}
\makeatletter
\AtBeginDocument{%
  \fancypagestyle{briefreview}[standardpagestyle]{%
    \fancyhead{}%
    \fancyhead[LO,LE]{\ACM@linecountL}%
  }%
  \fancypagestyle{brieffirst}[firstpagestyle]{%
    \fancyhead{}%
    \fancyhead[L]{\ACM@linecountL}%
  }%
}
\makeatother

\usepackage{xspace}
\usepackage{fancyvrb} %
\usepackage{tikz}
\usepackage{multirow}
\usepackage{adjustbox}
\usepackage{pgfplots}
\usepgfplotslibrary{groupplots}
\pgfplotsset{compat=1.17}
\usepackage[all]{hypcap}
\usepackage{placeins}
\usepackage[most]{tcolorbox}
\tcbuselibrary{breakable}
\newtcolorbox{findingbox}[1]{
  title={\textbf{Finding #1}},
  colback=white,
  colframe=gray!50,
  colbacktitle=gray!55,
  coltitle=black,
  fonttitle=\bfseries,
  boxrule=0.5pt,
  arc=2pt,
  titlerule=0pt,
  toptitle=2pt,
  bottomtitle=2pt,
  breakable
}
\newcounter{finding}
\newenvironment{finding}{%
  \refstepcounter{finding}%
  \begin{findingbox}{\thefinding}%
}{\end{findingbox}}

\newcommand{\brief}{\textsc{Brief}\xspace}

\newcommand{\circled}[1]{\tikz[baseline=(char.base)]{\node[shape=circle,draw,inner sep=1.2pt,font=\scriptsize] (char) {#1};}}

\title{\brief: An Evidence-Assessed Repair-Intent Interface for LLM-Based Program Repair}

\author{Yi Su}
\affiliation{%
  \institution{University of Massachusetts Amherst}
  \city{Amherst}
  \state{Massachusetts}
  \country{USA}}

\author{Zhenghang Zhao}
\affiliation{%
  \institution{Columbia University}
  \city{New York}
  \state{New York}
  \country{USA}}

\author{Wei Cheng}
\affiliation{%
  \institution{NEC Laboratories America, Inc.}
  \city{Princeton}
  \state{New Jersey}
  \country{USA}}

\author{Haifeng Chen}
\affiliation{%
  \institution{NEC Laboratories America, Inc.}
  \city{Princeton}
  \state{New Jersey}
  \country{USA}}

\author{Juan Zhai}
\email{juanzhai@umass.edu}
\affiliation{%
  \institution{University of Massachusetts Amherst}
  \city{Amherst}
  \state{Massachusetts}
  \country{USA}}

\authorsaddresses{Corresponding author: Juan Zhai,
\href{mailto:juanzhai@umass.edu}{juanzhai@umass.edu}, University of
Massachusetts Amherst, Amherst, Massachusetts, USA}

\begin{document}

\begin{abstract}
Large language models (LLMs) have substantially advanced automated program
repair (APR) by reasoning over code, tests, and execution feedback. Recent
approaches further externalize pre-repair reasoning as diagnoses, plans, or
repair hypotheses. However, such guidance is typically produced for and
consumed within the workflow that creates it, and the guidance itself may be
unsupported, incomplete, or inconsistent before patch generation begins.
This leaves an important gap: repair intent should be reusable across repair
backends and assessable before a backend acts on it.

We introduce \brief, an evidence-guided repair harness that externalizes repair
intent as a backend-decoupled, evidence-assessed interface between project
evidence and patch generation. \brief synthesizes natural-language
documentation, failing tests, and code context into a structured Repair Plan
that captures intended behavior, observed violations, repair requirements,
target locations, and repair direction. A Plan Assessor checks the plan against
its source evidence for grounding, target consistency, coverage, and
concreteness, and rejected plans are refined before use. Only an accepted plan
is supplied to the downstream backend; if no plan is accepted, the backend
proceeds with its original repair procedure.

We evaluate \brief on all 835 bugs from 17 Defects4J projects using two
repair backends, RepairAgent and ReinFix, and two LLMs, DeepSeek-V3.2 and
GPT-6-Luna. Across all four backend--model configurations, \brief increases
correct repairs by 117--197 bugs, corresponding to relative gains of
60.9\%--333.9\%, and correctly repairs 86--206 bugs per configuration for
which the corresponding baseline produces no plausible patch. Ablation results further show that structuring the same evidence into a Repair
Plan increases correct repairs from 127 to 216 (+70.1\%), while assessment and
refinement further increase them to 256 (+18.5\%). These gains do not require proportional resource overhead: average
per-bug cost decreases in three of four configurations, by as much as 60.1\%,
and repair iterations decrease in all four. These results support assessed
repair intent as an effective, backend-decoupled interface for LLM-based
program repair.
\end{abstract}

\ccsdesc[500]{Software and its engineering~Automatic programming}
\ccsdesc[500]{Software and its engineering~Software testing and debugging}
\ccsdesc[500]{Software and its engineering~Software maintenance tools}
\keywords{automated program repair, large language models, agentic program repair,
repair intent, repair planning}

\maketitle
\pagestyle{briefreview}
\thispagestyle{brieffirst}

\section{Introduction}
\label{sec:introduction}

Automated program repair (APR) aims to reduce the effort developers spend
diagnosing and fixing software defects. Traditional APR techniques include
search- and template-based approaches that explore candidate patches using
predefined or mined repair patterns~\cite{legoues2012genprog,liu2019tbar}, constraint-based approaches
that synthesize patches from symbolic correctness conditions~\cite{nguyen2013semfix,mechtaev2016angelix}, and
learning-based approaches that learn program transformations from historical
bug-fix data~\cite{chen2021sequencer,lutellier2020coconut}. More recently, large language models (LLMs) have
enabled repair systems that generate patches from buggy code and its context
~\cite{xia2023automated,jiang2023impact} and, increasingly, interact with
development tools, execution results, and test feedback to iteratively refine
candidate fixes~\cite{bouzenia2025repairagent,xia2024automated,kang2025explainable}.
These systems substantially broaden the information and actions available
during repair. Nevertheless, producing a correct patch still requires
interpreting what behavior the program is intended to implement, how the
current implementation violates that behavior, and which constraints a repair
must preserve.

The evidence needed for this interpretation is often available, but fragmented
across a project's natural-language (NL) documentation, tests, and code.
Documentation describes behavioral requirements and boundary conditions;
failing tests expose violations on concrete executions; code reveals how the
relevant behavior is implemented and how it interacts with surrounding
components. Each source provides a partial view of the defect. A failing
assertion identifies an unmet expectation but may leave the underlying
requirement implicit. A documented constraint states expected behavior but
must still be connected to the implementation that violates it. Code exposes
implementation details and candidate repair locations, but does not by itself
determine what behavioral obligations a correct change must satisfy. Together, these sources provide complementary evidence, but they do not by
themselves constitute actionable repair guidance.

LLM-based repair agents already perform substantial reasoning during this
process. They can inspect localized code, formulate hypotheses, retrieve
additional program context, generate candidate patches, and revise their
hypotheses after observing validation feedback
~\cite{bouzenia2025repairagent,zhang2025reinfix}. The challenge is therefore not
simply to provide an LLM with more information. Even when relevant evidence is
available, the backend must still reconcile behavioral requirements, observed
failures, and implementation details into a coherent repair intent. This
interpretation may itself be incomplete or incorrect: a model may misread a
documented constraint, overlook a failing scenario, or pursue a repair direction
that is not supported by the available evidence. When repair intent remains implicit within the backend's evolving repair
trajectory, its quality cannot be examined independently before it influences
patch generation.

The motivating example in Section~\ref{sec:motivation} illustrates this gap.
RepairAgent has access to the relevant code, tests, and execution feedback, yet
exhausts its interaction budget without producing a correct repair. When the
same backend is instead supplied with an explicit Repair Plan that connects the
available evidence to the required repair behavior, it successfully repairs the
defect. This case highlights the distinction between merely accessing repair evidence
and turning that evidence into explicit repair intent that can be examined
before patch generation.

Recent work has increasingly made pre-repair reasoning explicit.
PracRepair performs evidence-driven diagnosis and summarizes the resulting
analysis into repair hypotheses that guide subsequent patch generation
~\cite{cheng2026pracrepair}, while CAMP separates pre-repair reasoning from
repair and structures that reasoning around domain knowledge, intended
functionality, and bug causes~\cite{wu2026camp}. These approaches demonstrate
the value of deriving explicit repair guidance before generating a patch.
However, two limitations remain when such guidance is used to drive repair.
First, the resulting hypotheses or reasoning artifacts are primarily produced
and consumed within their respective end-to-end repair workflows, rather than
exposed as a reusable interface for existing repair backends. Second, explicit
guidance is itself a potential failure point: it may misinterpret a behavioral
requirement, omit a failing scenario, identify an inconsistent repair target,
or recommend a change unsupported by the available evidence. Such errors can
propagate directly into patch generation unless the guidance is checked before
it is used.

These limitations point to a missing abstraction in current LLM-based repair:
a repair-intent interface that is both \emph{backend-decoupled} and
\emph{evidence-assessed}. We introduce \brief, an evidence-guided repair harness
that externalizes repair intent as an explicit, evidence-assessed interface
between project evidence and patch generation. Given a repair task, \brief
collects NL documentation, failing tests, and code context into an
\emph{Evidence Bundle}. A Repair Plan Synthesizer converts this material into
a \emph{Candidate Repair Plan} that connects intended behavior, observed
violations, repair requirements, target locations, and repair direction. The
plan specifies what a successful repair should accomplish without prescribing
the backend's reasoning process or the concrete implementation of the patch.

Making repair intent explicit also makes it possible to assess that intent
before it influences patch generation. A dedicated Plan Assessor checks each
Candidate Repair Plan against the original evidence for \emph{grounding},
\emph{target consistency}, \emph{coverage}, and \emph{concreteness}. When a
candidate fails these checks, the assessor returns structured feedback and the
synthesizer revises the plan. Only an accepted plan is supplied to the repair
backend. This assessment--refinement process acts as a pre-generation quality
gate for detecting unsupported, incomplete, inconsistent, or imprecise repair
guidance before it is consumed downstream.

This design separates evidence interpretation and assessment from patch
generation. An existing backend can consume the Accepted Repair Plan as
additional structured context while retaining its native tools,
patch-generation strategy, validation procedure, and refinement loop. The plan
therefore serves as a stable interface between \brief's evidence analysis and
the backend's repair procedure. If no plan is accepted within the configured
assessment--refinement budget, \brief discards the candidate guidance and the
backend proceeds with its original inputs and procedure. Thus, \brief augments rather than replaces the backend: it determines whether
repair guidance is sufficiently supported to be used, while the backend retains
control over how the patch is generated and validated.

We evaluate \brief on all 835 bugs from 17 Defects4J projects
~\cite{just2014defects4j}, using two LLM-based repair backends,
RepairAgent~\cite{bouzenia2025repairagent} and
ReinFix~\cite{zhang2025reinfix}, with two LLMs,
DeepSeek-V3.2~\cite{deepseekai2025deepseekv32} and
GPT-6-Luna~\cite{openai2026gpt6luna}. For each backend--model
configuration, we compare the original backend with the same backend augmented
by \brief under the same model, repair budget, execution environment, and
repair-task inputs, thereby isolating the effect of adding \brief.

\brief improves correct repair effectiveness in all four configurations,
adding 117--197 correct repairs and yielding relative gains of
60.9\%--333.9\%. The gains extend beyond easy cases: they hold for both
single-hunk and multi-hunk bugs, and \brief correctly repairs 86--206 bugs per
configuration for which the corresponding baseline produces no plausible
patch. The improvement is also not explained by evidence injection alone: with
RepairAgent and GPT-6-Luna, direct evidence injection yields 127 correct
repairs, structuring the same evidence into a Repair Plan raises this to 216
(+70.1\%), and assessment and refinement further increase it to 256 (+18.5\%). Finally, these gains do not require proportional resource overhead:
average per-bug cost decreases in three of four configurations, by as much as
60.1\%, and repair iterations decrease in all four.

To assess the potential influence of training-data
contamination, we test \brief on 41 Java bugs from
SWE-bench-Live/MultiLang~\cite{zhang2025swebenchlive,li2026repolaunch}
created after DeepSeek-V3.2's release. With DeepSeek-V3.2, correct repairs
increase from 5 to 8 for RepairAgent and from 13 to 16 for ReinFix.
These results suggest that \brief's performance on Defects4J is not
much affected by potential data leakage.

Together, these results show that assessed repair intent increases
correct repairs across the four Defects4J configurations and on the
additional benchmark, while lowering average cost in three of the four
Defects4J configurations.

This paper makes three contributions:
\begin{itemize}
\item \textbf{A backend-decoupled repair-intent interface.}
We introduce \brief, which externalizes repair intent as a structured Repair
Plan connecting intended behavior, observed violations, repair requirements,
target locations, and repair direction. The plan specifies what a repair
should accomplish while leaving patch generation, validation, and refinement
to the existing repair backend, enabling the same interface to augment
different repair procedures without redesigning their native workflows.

\item \textbf{An evidence-assessed quality gate before patch generation.}
We introduce an assessment--refinement mechanism that checks Candidate Repair
Plans against their source evidence for grounding, target consistency,
coverage, and concreteness before they can influence patch generation.
Rejected plans are revised using structured feedback, and plans that remain
unaccepted are withheld from the backend, preventing unsupported repair
guidance from being imposed on downstream repair.

\item \textbf{Cross-backend empirical evidence for assessed repair intent.}
We evaluate \brief on all 835 Defects4J bugs using RepairAgent and ReinFix
with DeepSeek-V3.2 and GPT-6-Luna. Across all four backend--model
configurations, \brief increases correct repairs by 60.9\%--333.9\%, with
gains on both single-hunk and multi-hunk bugs. In the cumulative ablation,
structuring raw evidence into a Repair Plan increases correct repairs from
127 to 216 (+70.1\%), and assessment further increases them to 256 (+18.5\%).
Average per-bug cost also decreases in three of four configurations, while
repair iterations decrease in all four.
\end{itemize}

\section{Motivation}
\label{sec:motivation}

\begin{figure*}[!t]
    \centering
    \footnotesize
    \input{figs/moti-cli30}
\caption{Motivating example: fragmented repair evidence is
synthesized into an explicit Repair Plan that guides the backend to the correct
fix for Cli-30.}
\label{fig:moti}
\end{figure*}

Figure~\ref{fig:moti} shows Cli-30, a Defects4J bug in Apache Commons CLI's
property-based option handling. The defect occurs in
\texttt{DefaultParser.handleProperties()}, where a property may select an
option from a group whose another member has already been selected. In our
observed run, RepairAgent~\cite{bouzenia2025repairagent} exhausts its interaction
budget without producing a correct patch. When supplied with the Repair Plan
synthesized by \brief, however, the same backend successfully repairs the
defect. This contrast highlights the gap between accessing repair evidence and
turning that evidence into explicit, actionable repair intent.

Figure~\ref{fig:moti}(a) shows the buggy method
\texttt{handleProperties()} in \texttt{DefaultParser.java}. The inline
documentation at lines~153--154 explicitly states that, if an option belongs
to a group, the parser should check whether another member of that group has
already been selected. However, the implementation contains no corresponding
check before the property option is processed. The failing test in
Figure~\ref{fig:moti}(b) makes this requirement concrete. The command-line
arguments select option \texttt{a}, while the test supplies property
\texttt{b}, which belongs to the same option group. The call at line~1037 then
produces the failure trace in Figure~\ref{fig:moti}(c). Lines~1--2 of the trace
report an \texttt{AlreadySelectedException}: option \texttt{b} is being
selected even though option \texttt{a} from the same group has already been
selected. The subsequent frames show that the conflicting option propagates
through \texttt{handleProperties()} and \texttt{handleOption()} before the
exception is raised.

From this distributed evidence, \brief synthesizes the Repair Plan shown in
Figure~\ref{fig:moti}(d). The \emph{intended purpose} field
(\circled{1}) summarizes the desired behavior: property processing should
preserve an already selected option group. The \emph{documentation constraints}
field (\circled{2}) records the requirement expressed by the inline comment,
while the \emph{code violations} field (\circled{3}) identifies the missing
group-conflict check in \texttt{handleProperties()} and explains how the
conflicting option can reach \texttt{handleOption()}. The
\emph{test expectation} field (\circled{4}) captures the required behavior of
the failing scenario, and the \emph{rules} field (\circled{5}) turns these
observations into an explicit constraint that a property must not select a
second member of an already selected option group. Finally, the
\emph{fix target} (\circled{6}) points to the omission after the option lookup,
and the \emph{fix direction} (\circled{7}) specifies that the missing
group-selection logic should be added.

The resulting plan therefore makes explicit both \emph{where} the repair is
needed and \emph{what} behavioral constraint the repair must enforce, while
leaving the concrete implementation to the repair backend. With this guidance, RepairAgent successfully generates the correct fix for
Cli-30, demonstrating the value of supplying an explicit repair intent to the
backend before patch generation.

This contrast motivates \brief: repair evidence may already be available, yet
remain fragmented across documentation, tests, execution traces, and code.
\brief consolidates these fragments into an explicit Repair Plan before patch
generation.

\section{Methodology}
\label{sec:methodology}

\begin{figure*}[t]
    \centering
    \footnotesize
    \includegraphics[
    width=\textwidth,
    ]{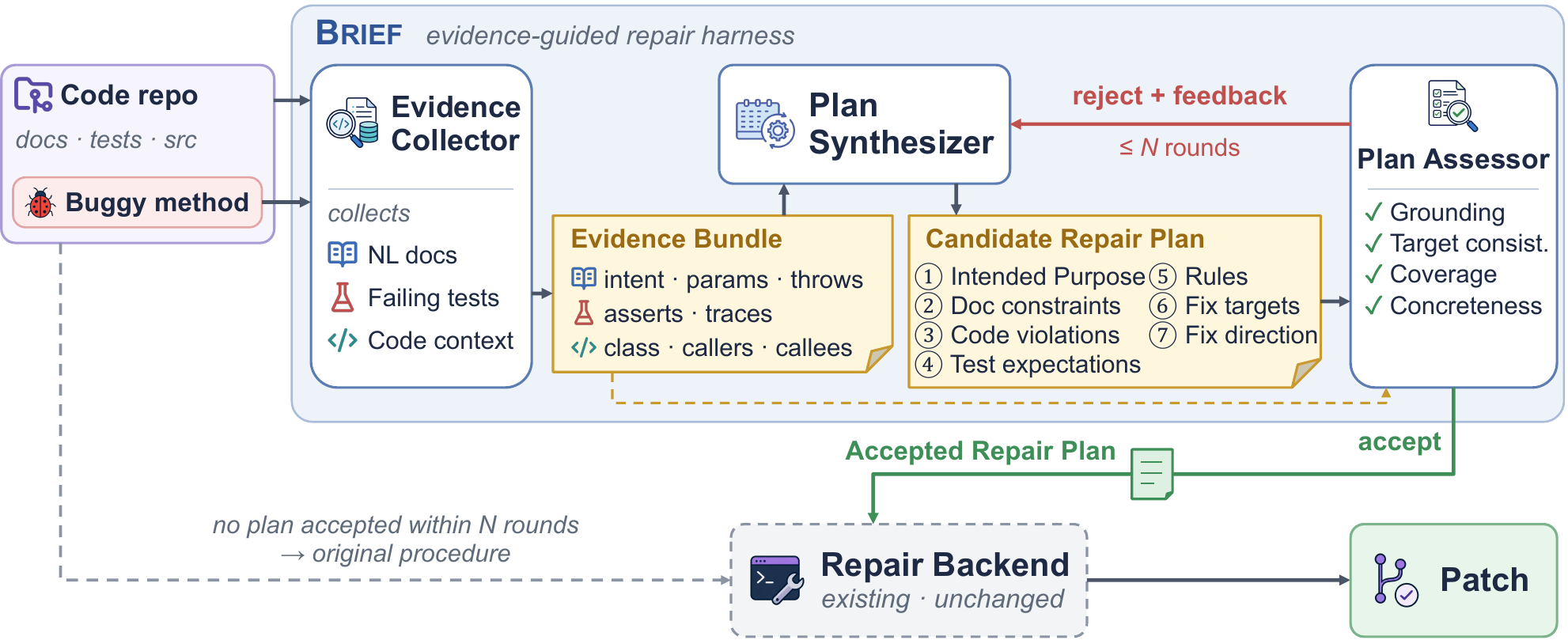}
\caption{Overview of \brief. Project evidence is synthesized into a Repair Plan,
assessed and refined before use, and supplied to an unchanged repair backend
only after acceptance.}
\label{fig:overview}
\end{figure*}

\subsection{Overview}
\brief is an evidence-guided repair harness that sits between project evidence
and an existing repair backend, as illustrated in Figure~\ref{fig:overview}.
Given a code repository and a repair target, the Evidence Collector
gathers three complementary sources of information: NL documentation, failing
tests, and code context. These artifacts are organized into an
\textit{Evidence Bundle}, which preserves the behavioral requirements,
observed failures, and implementation context relevant to the repair task.

The Repair Plan Synthesizer interprets this bundle and produces a
\textit{Candidate Repair Plan}. The plan makes the repair intent explicit by
recording the intended purpose, documentation constraints, code violations,
test expectations, repair rules, fix targets, and fix direction. Before this
guidance can influence patch generation, the Plan Assessor checks the candidate
against the original evidence for grounding, target consistency, coverage, and
concreteness. A rejected plan is returned to the synthesizer together with
assessment feedback and revised for another round. This assessment--refinement
loop continues for at most \(N\) rounds, where \(N\) is a configurable
refinement budget.

Only an accepted candidate becomes an \textit{Accepted Repair Plan}. \brief
supplies this plan to the downstream repair backend as structured context,
while leaving the backend's native repair procedure unchanged. The backend
remains responsible for patch generation, validation, and iterative refinement.
If no plan is accepted within \(N\) rounds, \brief discards the candidate
guidance and the backend proceeds using its original inputs and procedure.

This architecture separates repair-intent synthesis and assessment from patch
generation. The Evidence Bundle represents the collected project evidence, the
Repair Plan represents an assessed interpretation of what the repair should
accomplish, and the backend determines how to realize that intent as an
executable patch. In this way, \brief provides an explicit, evidence-assessed
interface between project evidence and existing repair procedures.

\subsection{Evidence Collector}
\label{sec:evidence}

The Evidence Collector gathers information about a buggy method from NL documentation, failing tests, and code context. These sources describe intended behavior, observed failures, and implementation context. Each source can be incomplete or ambiguous. The collector combines them into an \textit{Evidence Bundle} for the Repair Plan Synthesizer.

\textbf{NL documentation} states the method's intended behavior and constraints. \brief extracts method summaries, parameter constraints, return-value guarantees, exceptional behavior, and boundary cases from structured tags and prose. These statements can describe requirements that a particular failing execution does not expose.
Here, NL documentation includes method comments such as those in Figure~\ref{fig:moti}; it need not be a separate external artifact. 

\textbf{Failing tests} show how the implementation violates expected behavior. For each test, \brief collects the test class and method name, full source, and failure output. The output includes assertion failures, observed values, exception messages, and stack traces. This evidence connects the repair task to concrete inputs, outputs, and execution locations. A stack-trace location identifies an observed execution point; it does not by itself establish a faulty line.

\textbf{Code context.}
\brief uses lightweight source-level extraction to assemble bounded
implementation context around the designated repair target. The collector
retrieves the enclosing method and class, same-name sibling methods, and
selected usage context through lexical matching and brace-based extraction.
These artifacts provide the synthesizer with local implementation structure and
additional surrounding context that may help interpret the repair target. All
retrieved context is truncated to predefined size limits before being included
in the Evidence Bundle.

\subsection{Repair Plan Synthesizer}
\label{sec:generation}

The Repair Plan Synthesizer converts the Evidence Bundle into a \textit{Candidate Repair Plan}. It identifies the expected behavior, explains how the implementation violates that behavior, and specifies the required locations and direction of change. The plan provides a compact interface between project evidence and downstream repair. The Plan Assessor checks this candidate before the backend uses it to generate a patch.

The Candidate Repair Plan has seven components, shown in Figure~\ref{fig:moti}(d).
\circled{1}~\textbf{Intended purpose} states the method's behavioral goal based on documentation and code context.
\circled{2}~\textbf{Documentation constraints} records the documented obligations relevant to repair and omits descriptions that do not constrain behavior.
\circled{3}~\textbf{Code violations} identifies locations where the implementation conflicts with those obligations and describes each discrepancy.
\circled{4}~\textbf{Test expectations} states the expected behavior for each failing test, based on its logic and observed failure.
\circled{5}~\textbf{Rules} combines constraints from documentation, tests, and code into conditions the repair must satisfy.
\circled{6}~\textbf{Fix targets} identifies the affected locations and the required behavioral change at each location.
\circled{7}~\textbf{Fix direction} specifies whether the repair requires adding, removing, or modifying code.
An explicit edit direction helps counter the additive bias documented in LLMs on general modification tasks~\cite{santagata2024more}, which may lead models to add code even when a repair requires deletion.

\subsection{Repair Plan Assessment and Refinement}
\label{sec:auditor}

The Repair Plan Assessment and Refinement component forms a quality-control
loop between plan synthesis and patch generation. Its goal is to prevent
unsupported or incomplete repair intent from being passed directly to the
repair backend and, when possible, to correct such deficiencies before the
backend acts on the plan. A Candidate Repair Plan may appear plausible while
still misinterpreting a behavioral requirement, omitting a failing scenario,
identifying a target inconsistent with the available program context, or
providing guidance that is too vague to implement. Rather than immediately
using or discarding such a plan, \brief checks it against the evidence from
which it was synthesized and uses the resulting feedback to refine the plan.

The Plan Assessor receives the Candidate Repair Plan together with the original
Evidence Bundle. It uses the same LLM as the synthesizer in a separate
assessment invocation and does not receive the synthesizer's reasoning trace.
The assessor therefore evaluates the resulting guidance against its source
evidence rather than the reasoning that produced it. The separation does not
guarantee correctness: the synthesizer and assessor may share model biases or
misinterpret the same evidence. An \textit{Accepted Repair Plan} therefore
indicates that the plan satisfies the specified assessment criteria, not that
the diagnosis or eventual patch is guaranteed to be correct.

The assessor evaluates each Candidate Repair Plan along four dimensions:
\textit{grounding}, \textit{target consistency}, \textit{coverage}, and
\textit{concreteness}. These dimensions are design criteria for assessing
repair guidance, rather than an empirically derived taxonomy of failures in
LLM-based repair.

\textbf{Grounding} checks whether behavioral requirements, violation claims,
and repair directions are supported by identifiable evidence, such as
documentation, test assertions, execution outputs, or code observations.
Unsupported claims are flagged for revision.

\textbf{Target consistency} checks whether each proposed repair target and its
described violation are consistent with the supplied program context. This
criterion checks how the plan uses available localization information; it does
not measure independent fault-localization ability.

\textbf{Coverage} checks whether the plan accounts for the relevant
requirements and discrepancies exposed by the collected evidence. The assessor
flags failing scenarios or supported discrepancies that are omitted from the
repair guidance.

\textbf{Concreteness} checks whether the plan provides sufficiently actionable
guidance for downstream repair. The assessor flags unresolved alternatives,
ambiguous directions, and repair decisions that remain unspecified.

For each Candidate Repair Plan, the assessor returns either an acceptance
decision or structured feedback describing the detected deficiencies. An
accepted candidate becomes the \textit{Accepted Repair Plan} and is supplied
to the downstream repair backend. For a rejected candidate, the Repair Plan
Synthesizer revises the plan using both the assessor feedback and the original
Evidence Bundle, after which the revised candidate is assessed again. This
process forms an assessment--refinement loop that improves repair guidance
before it reaches patch generation.

\brief allows at most three assessment--refinement rounds in our main
experiments. We selected this budget based on a six-round pilot study with
DeepSeek-V3.2, where
Round~3 still yielded accepted plans for 10.0\% of the sampled
bugs, compared with only 1.7\%, 1.7\%, and 0\% in Rounds~4--6.
Section~\ref{sec:assess_experiment} reports the round-wise plan acceptance
and repair outcomes under this three-round budget.

If no candidate is accepted within the three-round budget, \brief discards the
candidate plan and falls back to the original backend without providing repair
guidance. The backend then proceeds with its native repair procedure and original
inputs. This fallback preserves the separation between \brief and the repair
backend: an unaccepted plan never constrains patch generation, and failure to
produce an accepted plan does not prevent the backend from attempting the repair
independently.

\begin{figure*}[!t]
    \centering
    \footnotesize
    \includegraphics[
    width=.999\textwidth,
    ]{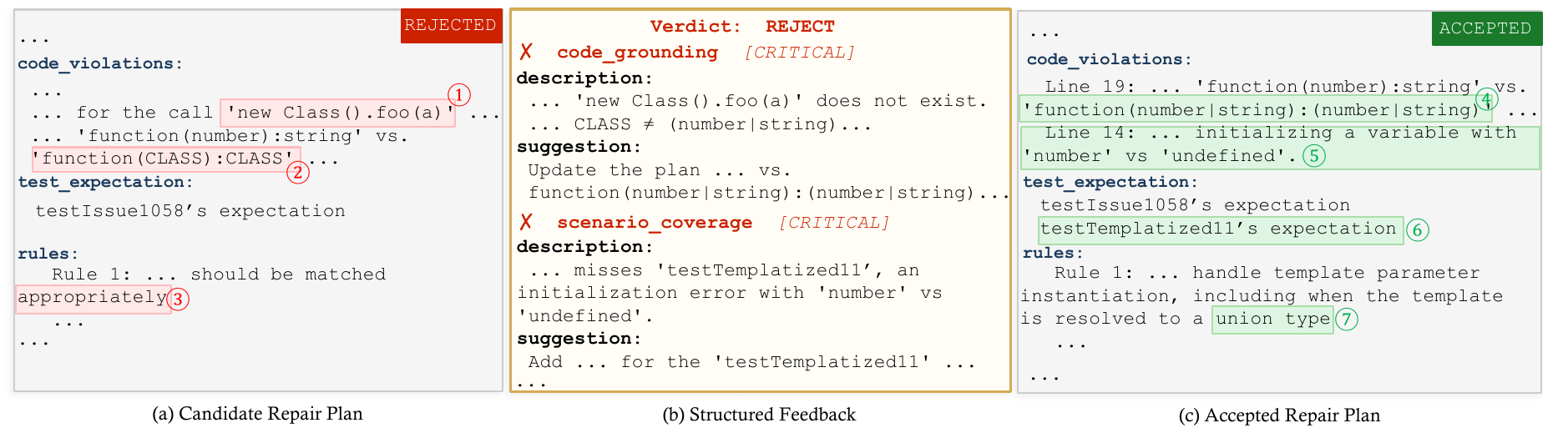}
    \caption{Plan assessment and refinement for Closure-112.
    Some fields are omitted for brevity.}
    \label{fig:audit}
\end{figure*}

\paragraph{Illustrative example.}
Figure~\ref{fig:audit} uses Closure-112, a type-checking bug in the Closure
Compiler, to illustrate how \brief assesses and refines a Candidate Repair Plan.
Figures~\ref{fig:audit}(a)--(c) show the initial plan, the assessor feedback,
and the accepted revision, respectively.

The assessor first identifies two grounding problems. The
\texttt{code\_violations} field refers to
\texttt{new Class().foo(a)}~(\circled{1}), which does not appear in the source;
the synthesizer therefore removes this reference. The candidate also states the
required type as \texttt{function(CLASS):CLASS}~(\circled{2}), whereas the
failing-test output reports
\texttt{function((number|string)):(number|string)}. The revised plan corrects
this type~(\circled{4}).

The assessor also identifies a coverage problem. The
\texttt{test\_\allowbreak{}expectation} field represents only one of the two
failing tests and omits \texttt{testTemplatized11}. In response, the
synthesizer adds the missing expectation~(\circled{6}) and the corresponding
\texttt{code\_violations} entry~(\circled{5}). The revised plan then passes
assessment and becomes the Accepted Repair Plan supplied to the backend.

\subsection{Integration with Repair Backends}
\label{sec:repair}

Once a Candidate Repair Plan passes assessment, \brief supplies the resulting
\textit{Accepted Repair Plan} to the downstream LLM-based repair backend as
structured context. The plan summarizes the intended behavior, observed
violations, repair requirements, target locations, and repair direction. It
specifies what the repair should accomplish without prescribing the concrete
patch or the backend's reasoning procedure.

The backend consumes the plan together with its native repair context and
retains responsibility for patch generation, validation, and refinement. For
iterative backends, the Accepted Repair Plan remains available throughout the
repair session as a stable statement of repair intent while the backend manages
patch attempts, test results, and repair history. Because evidence collection,
plan synthesis, and assessment occur outside the backend, integration requires
only that the backend accept additional structured context; its native repair
procedure remains unchanged. The Accepted Repair Plan therefore serves as the
interface between \brief's evidence analysis and the backend's patch-generation
process.

\section{Experimental Setup}
\label{sec:experimental-setup}

We evaluate \brief by addressing the following three research questions:

\noindent\textbf{RQ1 (Repair Effectiveness).}
\emph{How effectively does \brief improve repair performance across different repair backends and LLMs?}

\noindent\textbf{RQ2 (Design Analysis).}
\emph{How do \brief's evidence synthesis and plan assessment contribute to its repair effectiveness?}

\noindent\textbf{RQ3 (Repair Efficiency).}
\emph{How does augmenting repair backends with \brief affect their computational cost and repair efficiency?}

\subsection{Benchmark}

We evaluate \brief on Defects4J~\cite{just2014defects4j}, a widely used benchmark for automated program repair research. We use Defects4J v2.0, which contains 835 active real-world bugs from 17 Java projects: the 6 projects of Defects4J v1.2 and 11 additional projects introduced in v2.0. Each bug is accompanied by triggering tests and a developer patch, providing reproducible buggy program versions and a reference for assessing patch correctness. To assess the potential influence of training-data contamination in our Defects4J results, we additionally evaluate both repair backends with DeepSeek-V3.2 on 41 Java bugs from 11 projects in SWE-bench-Live/MultiLang~\cite{zhang2025swebenchlive,li2026repolaunch,swebenchlivemultilangdataset}. All 41 bugs were created after DeepSeek-V3.2's December 1, 2025 release. We use only DeepSeek-V3.2 in this evaluation because, at the time of our study, no benchmark of Java bugs created after GPT-6-Luna's release was yet available.

\subsection{Repair Backends and LLMs}

We evaluate \brief with two LLM-based repair backends,
RepairAgent~\cite{bouzenia2025repairagent} and
ReinFix~\cite{zhang2025reinfix}, and two LLMs,
DeepSeek-V3.2~\cite{deepseekai2025deepseekv32} and
GPT-6-Luna~\cite{openai2026gpt6luna}. For each backend--model configuration,
we compare the original backend with the same backend augmented by \brief.
The paired configurations use the same model, repair budget, execution
environment, and repair inputs; the only difference is that the augmented
configuration receives an Accepted Repair Plan before patch generation.

\subsection{Repair Success Criteria}

A patch is \emph{plausible} if the project compiles and passes all available developer tests. A plausible patch may still overfit to those tests. A patch is \emph{correct} if it implements the intended behavior. We first compare plausible patches with the developer patch and classify syntactically equivalent patches as correct. We then manually inspect the remaining patches against the developer patch and intended behavior to assess semantic correctness.

\section{Results}
\label{sec:results}

\subsection{RQ1: Repair Effectiveness}
\label{sec:rq1}
\subsubsection{Overall Repair Effectiveness}

\providecommand{\tool}{\textsc{Brief}}
\providecommand{\gptmodel}{GPT-6-Luna}
\providecommand{\backendB}{ReinFix}

\begin{table}[t]
\centering
\caption{Overall repair performance on Defects4J.}
\label{tab:full-benchmark-results}
\small
\begin{adjustbox}{max width=\linewidth}
\begin{tabular}{lllrrr}
\toprule
Backend & Model & Configuration & \#Plausible & \#Correct & $\Delta$\#Correct \\
\midrule
\multirow{4}{*}{RepairAgent} & \multirow{2}{*}{DeepSeek-V3.2} & Baseline & 77 & 66 & -- \\
 & & +\tool & \textbf{238} & \textbf{204} & +138 (+209.1\%) \\
\cmidrule{2-6}
 & \multirow{2}{*}{\gptmodel} & Baseline & 74 & 59 & -- \\
 & & +\tool & \textbf{336} & \textbf{256} & +197 (+333.9\%) \\
\midrule
\multirow{4}{*}{\backendB} & \multirow{2}{*}{DeepSeek-V3.2} & Baseline & 288 & 192 & -- \\
 & & +\tool & \textbf{385} & \textbf{309} & +117 (+60.9\%) \\
\cmidrule{2-6}
 & \multirow{2}{*}{\gptmodel} & Baseline & 293 & 203 & -- \\
 & & +\tool & \textbf{407} & \textbf{327} & +124 (+61.1\%) \\
\bottomrule
\end{tabular}
\end{adjustbox}
\end{table}

Table~\ref{tab:full-benchmark-results} compares each repair backend's
original version (Baseline) with its +\brief variant under two language
models on 835 Defects4J bugs. $\Delta$\#Correct is the increase in correct
repairs over the corresponding baseline.

With RepairAgent, +\brief raises plausible repairs from 77 to 238 and
correct repairs from 66 to 204 under DeepSeek-V3.2 (+138 correct repairs,
+209.1\%). Under \gptmodel{}, the counts rise from 74 to 336 plausible
repairs and from 59 to 256 correct repairs (+197, +333.9\%). With
\backendB{}, plausible repairs rise from 288 to 385 and correct repairs
from 192 to 309 under DeepSeek-V3.2 (+117, +60.9\%); under \gptmodel{},
they rise from 293 to 407 and from 203 to 327, respectively (+124,
+61.1\%). The largest gain in correct repairs, both in count and
percentage, occurs with RepairAgent and \gptmodel{}.

\subsubsection{Patch Complexity and Repair Coverage}

\providecommand{\tool}{\textsc{Brief}}
\providecommand{\gptmodel}{GPT-6-Luna}
\providecommand{\backendB}{ReinFix}

\definecolor{baseorange}{RGB}{244,164,96}
\definecolor{oursblue}{RGB}{70,130,180}
\pgfplotsset{
  evalchart/.style={
    ybar=0pt, clip=false,
    scale only axis, width=0.205\textwidth, height=3.15cm,
    ymin=0,
    ymajorgrids, grid style={dashed, gray!35},
    axis lines*=left, axis line style={gray!70},
    xtick style={draw=none},
    tick label style={font=\footnotesize}, title style={font=\small, yshift=-2pt},
    ylabel style={font=\small},
    unbounded coords=discard,
    nodes near coords, nodes near coords style={font=\scriptsize},
    legend style={draw=none, font=\small, legend columns=2,
                  /tikz/every even column/.append style={column sep=14pt}},
    area legend,
  },
  hunkchart/.style={evalchart, bar width=7pt, ymax=350, ytick distance=50,
    symbolic x coords={Single-hunk,Multi-hunk}, xmin=Single-hunk, xmax=Multi-hunk,
    xtick={Single-hunk,Multi-hunk},
    xticklabels={{Single-\\hunk},{Multi-\\hunk}},
    xticklabel style={align=center}, enlarge x limits=0.5},
  basebar/.style={fill=baseorange, draw=none,
                  nodes near coords style={font=\scriptsize, text=gray!60!black}},
  oursbar/.style={fill=oursblue, draw=none,
                  nodes near coords style={font=\scriptsize\bfseries, text=oursblue}},
}
\tikzset{gainlabel/.style={font=\scriptsize\bfseries, text=oursblue,
                           anchor=south, xshift=4pt, yshift=13pt}}

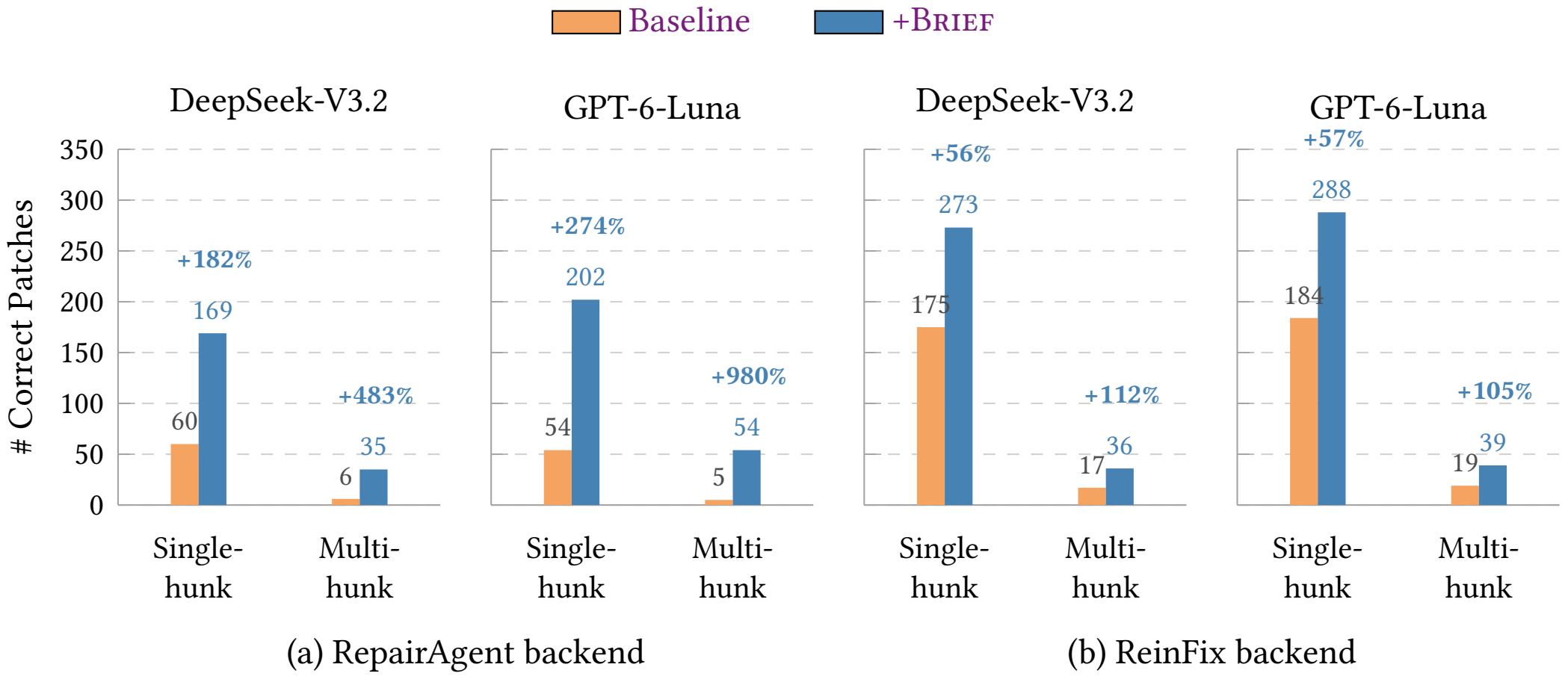
\begin{figure*}[t]
\centering
\ref{hunklegend}\\[4pt]
\begin{adjustbox}{max width=\linewidth}
\begin{tikzpicture}
\begin{groupplot}[group style={group size=4 by 1, horizontal sep=0.45cm,
                              y descriptions at=edge left}, hunkchart]
\nextgroupplot[title={DeepSeek-V3.2}, ylabel={\# Correct Patches}, legend to name=hunklegend]
\addplot[basebar] coordinates {(Single-hunk,60) (Multi-hunk,6)};
\addplot[oursbar] coordinates {(Single-hunk,169) (Multi-hunk,35)};
\node[gainlabel] at (axis cs:Single-hunk,169) {+182\%};
\node[gainlabel] at (axis cs:Multi-hunk,35) {+483\%};
\legend{Baseline, +\tool}
\nextgroupplot[title={\gptmodel}]
\addplot[basebar] coordinates {(Single-hunk,54) (Multi-hunk,5)};
\addplot[oursbar] coordinates {(Single-hunk,202) (Multi-hunk,54)};
\node[gainlabel] at (axis cs:Single-hunk,202) {+274\%};
\node[gainlabel] at (axis cs:Multi-hunk,54) {+980\%};
\nextgroupplot[title={DeepSeek-V3.2}]
\addplot[basebar] coordinates {(Single-hunk,175) (Multi-hunk,17)};
\addplot[oursbar] coordinates {(Single-hunk,273) (Multi-hunk,36)};
\node[gainlabel] at (axis cs:Single-hunk,273) {+56\%};
\node[gainlabel] at (axis cs:Multi-hunk,36) {+112\%};
\nextgroupplot[title={\gptmodel}]
\addplot[basebar] coordinates {(Single-hunk,184) (Multi-hunk,19)};
\addplot[oursbar] coordinates {(Single-hunk,288) (Multi-hunk,39)};
\node[gainlabel] at (axis cs:Single-hunk,288) {+57\%};
\node[gainlabel] at (axis cs:Multi-hunk,39) {+105\%};
\end{groupplot}
\path (group c1r1.south west) --
  node[midway, below=30pt, font=\small] {(a) RepairAgent backend}
  (group c2r1.south east);
\path (group c3r1.south west) --
  node[midway, below=30pt, font=\small] {(b) \backendB{} backend}
  (group c4r1.south east);
\end{tikzpicture}
\end{adjustbox}
\caption{Correct patches by patch complexity with RepairAgent (a, left pair) and \backendB{} (b, right pair). Within each backend, DeepSeek-V3.2 is on the left and \gptmodel{} is on the right.}
\label{fig:hunk-bar}
\end{figure*}

Figure~\ref{fig:hunk-bar} breaks down correct repairs by patch complexity
over the full benchmark. +\brief improves both single-hunk and multi-hunk
repairs in every backend--model configuration. With RepairAgent under
DeepSeek-V3.2, correct repairs increase from 60 to 169 for single-hunk bugs
(+182\%) and from 6 to 35 for multi-hunk bugs (+483\%). Under GPT-6-Luna,
the corresponding increases are from 54 to 202 (+274\%) and from 5 to 54
(+980\%). The same pattern holds with ReinFix: under DeepSeek-V3.2,
single-hunk repairs increase from 175 to 273 (+56\%) and multi-hunk repairs
from 17 to 36 (+112\%); under GPT-6-Luna, they increase from 184 to 288
(+57\%) and from 19 to 39 (+105\%). Thus, despite variation in effect size
across backends and models, +\brief consistently improves both complexity
classes, with larger relative gains on multi-hunk bugs in all four
configurations.

\providecommand{\tool}{\textsc{Brief}}
\providecommand{\gptmodel}{GPT-6-Luna}
\providecommand{\backendB}{ReinFix}

\begin{table}[t]
\centering
\caption{Repair results on bugs for which the corresponding baseline produces
no plausible patch.}
\label{tab:unique}
\small
\begin{adjustbox}{max width=\linewidth}
\begin{tabular}{ll|rrr|rrr|rrr}
\toprule
\multirow{2}{*}{Backend} & \multirow{2}{*}{Model} & \multicolumn{3}{c|}{All} & \multicolumn{3}{c|}{Single-hunk} & \multicolumn{3}{c}{Multi-hunk} \\
 & & \#Bugs & \#Plausible & \#Correct & \#Bugs & \#Plausible & \#Correct & \#Bugs & \#Plausible & \#Correct \\
\midrule
\multirow{2}{*}{RepairAgent} & DeepSeek-V3.2 & 758 & 169 (22.3\%) & 140 (18.5\%) & 399 & 133 (33.3\%) & 113 (28.3\%) & 359 & 36 (10.0\%) & 27 (7.5\%) \\
 & \gptmodel & 761 & 279 (36.7\%) & 206 (27.1\%) & 397 & 200 (50.4\%) & 157 (39.5\%) & 364 & 79 (21.7\%) & 49 (13.5\%) \\
\midrule
\multirow{2}{*}{\backendB} & DeepSeek-V3.2 & 547 & 106 (19.4\%) & 86 (15.7\%) & 204 & 83 (40.7\%) & 67 (32.8\%) & 343 & 23 (6.7\%) & 19 (5.5\%) \\
 & \gptmodel & 542 & 116 (21.4\%) & 92 (17.0\%) & 201 & 89 (44.3\%) & 71 (35.3\%) & 341 & 27 (7.9\%) & 21 (6.2\%) \\
\bottomrule
\end{tabular}
\end{adjustbox}
\end{table}

Table~\ref{tab:unique} examines a stricter subset: bugs for which
the corresponding baseline produces no plausible patch. This analysis tests
whether \brief extends the observed repair coverage of the original backend.
With RepairAgent, +\brief correctly repairs 140 of 758 such bugs (18.5\%)
under DeepSeek-V3.2 and 206 of 761 (27.1\%) under GPT-6-Luna. With ReinFix,
it correctly repairs 86 of 547 (15.7\%) and 92 of 542 (17.0\%),
respectively.

Within this baseline-failure subset, multi-hunk bugs remain harder to recover.
In all four configurations, +\brief achieves a higher correct-repair rate on
single-hunk than on multi-hunk bugs. For example, with RepairAgent and
GPT-6-Luna, it correctly repairs 39.5\% of single-hunk failures (157/397),
compared with 13.5\% of multi-hunk failures (49/364). Taken together,
Figure~\ref{fig:hunk-bar} and Table~\ref{tab:unique} show that
\brief improves repair effectiveness across both complexity classes while
also recovering 86--206 bugs per configuration for which the corresponding
baseline produces no plausible patch.

\subsubsection{Effectiveness across projects}

\providecommand{\tool}{\textsc{Brief}}
\providecommand{\gptmodel}{GPT-6-Luna}

\begin{table*}[t]
\centering
\caption{Per-project correct repairs with RepairAgent.
$\Delta$ denotes +\brief minus the corresponding baseline.}
\label{tab:per_project_results}
\small
\begin{adjustbox}{max width=\linewidth}
\begin{tabular}{lr|rrr|rrr}
\toprule
\multirow{2}{*}{Project} & \multirow{2}{*}{\#Bugs} & \multicolumn{3}{c|}{DeepSeek-V3.2} & \multicolumn{3}{c}{\gptmodel} \\
 & & Baseline & +\tool & $\Delta$\#Correct & Baseline & +\tool & $\Delta$\#Correct \\
\midrule
Chart & 26 & 4 & 13 & +9 & 3 & 10 & +7 \\
Cli & 39 & 3 & 11 & +8 & 0 & 18 & +18 \\
Closure & 174 & 12 & 33 & +21 & 10 & 33 & +23 \\
Codec & 18 & 4 & 9 & +5 & 3 & 10 & +7 \\
Collections & 4 & 1 & 2 & +1 & 1 & 2 & +1 \\
Compress & 47 & 4 & 17 & +13 & 3 & 14 & +11 \\
Csv & 16 & 3 & 8 & +5 & 2 & 10 & +8 \\
Gson & 18 & 1 & 4 & +3 & 1 & 12 & +11 \\
JacksonCore & 26 & 2 & 6 & +4 & 0 & 9 & +9 \\
JacksonDatabind & 112 & 5 & 14 & +9 & 8 & 29 & +21 \\
JacksonXml & 6 & 0 & 0 & 0 & 0 & 1 & +1 \\
Jsoup & 93 & 6 & 21 & +15 & 8 & 31 & +23 \\
JxPath & 22 & 0 & 1 & +1 & 1 & 4 & +3 \\
Lang & 64 & 6 & 29 & +23 & 10 & 32 & +22 \\
Math & 106 & 11 & 25 & +14 & 4 & 29 & +25 \\
Mockito & 38 & 3 & 7 & +4 & 3 & 8 & +5 \\
Time & 26 & 1 & 4 & +3 & 2 & 4 & +2 \\
\midrule
Total & 835 & 66 & 204 & +138 & 59 & 256 & +197 \\
\bottomrule
\end{tabular}
\end{adjustbox}
\end{table*}

Table~\ref{tab:per_project_results} breaks down RepairAgent's correct
repairs by project and model, showing how +\brief's gains are distributed
across Defects4J.

Under DeepSeek-V3.2, +\brief increases correct repairs in 16 of 17 projects
and ties the baseline on JacksonXml. The largest absolute gains occur on Lang
(+23) and Closure (+21), contributing to an overall increase from 66 to 204
correct repairs (+138). Under GPT-6-Luna, +\brief increases correct repairs
in all 17 projects, with the largest gains on Math (+25), Closure (+23), and
Jsoup (+23); the overall count rises from 59 to 256 (+197). No project shows
a decrease under either model. These project-level results indicate that
\brief's aggregate gains are broadly distributed across the benchmark rather
than driven by a single project or a small subset of projects.

\subsubsection{Evaluation on Bugs Created After Model Release}
\label{sec:post-cutoff}

\providecommand{\tool}{\textsc{Brief}}
\providecommand{\backendB}{ReinFix}

\begin{table}[t]
\centering
\caption{Results on 41 SWE-bench-Live/MultiLang Java bugs (DeepSeek-V3.2).}
\label{tab:post-cutoff-results}
\small
\begin{adjustbox}{max width=\linewidth}
\begin{tabular}{lr|rrrr|rrrr}
\toprule
\multirow{3}{*}{Bug type} & \multirow{3}{*}{Bugs} & \multicolumn{4}{c|}{RepairAgent} & \multicolumn{4}{c}{\backendB} \\
 & & \multicolumn{2}{c}{Baseline} & \multicolumn{2}{c|}{+\tool} & \multicolumn{2}{c}{Baseline} & \multicolumn{2}{c}{+\tool} \\
 & & \#Plausible & \#Correct & \#Plausible & \#Correct & \#Plausible & \#Correct & \#Plausible & \#Correct \\
\midrule
Single-line & 7 & 4 & 4 & \textbf{5} & 4 & 2 & 1 & \textbf{4} & \textbf{3} \\
Multi-line & 23 & 3 & 1 & \textbf{5} & \textbf{3} & 12 & 10 & \textbf{13} & \textbf{11} \\
Multi-file & 11 & 1 & 0 & \textbf{2} & \textbf{1} & 4 & 2 & 4 & 2 \\
\midrule
Total & 41 & 8 & 5 & \textbf{12} & \textbf{8} & 18 & 13 & \textbf{21} & \textbf{16} \\
\bottomrule
\end{tabular}
\end{adjustbox}
\end{table}

Table~\ref{tab:post-cutoff-results} reports results on the 41
SWE-bench-Live/MultiLang Java bugs using DeepSeek-V3.2. With RepairAgent,
+\brief increases plausible repairs from 8 to 12 and correct repairs
from 5 to 8; with ReinFix, plausible repairs increase from 18 to 21
and correct repairs from 13 to 16. The additional correct repairs come
from multi-line and multi-file bugs with RepairAgent, and from single-line
and multi-line bugs with ReinFix. Because these bugs were created after
DeepSeek-V3.2's release, their specific bug-fix pairs could not have
appeared in the model's pre-release training data. This temporal separation
reduces concern about memorization of these fixes, although it does not
rule out other forms of leakage.

\begin{finding}
Across all four evaluated backend--model configurations, \brief increases
correct repairs by 117--197 bugs, corresponding to relative gains of
60.9\%--333.9\%. The gains hold for both single-hunk and multi-hunk bugs,
and \brief additionally recovers 86--206 bugs per configuration for which
the corresponding baseline produces no plausible patch. With RepairAgent,
the gains are distributed across 16 of 17 projects under DeepSeek-V3.2 and
all 17 projects under GPT-6-Luna. Together, these results show that \brief's
improvements are consistent across the evaluated backends, models, patch
complexities, and projects rather than being driven by a narrow subset of
cases. Results on bugs created after DeepSeek-V3.2's release further suggest that
\brief's Defects4J gains are not much affected by potential training-data
leakage.
\end{finding}
\FloatBarrier

\subsection{RQ2: Design Analysis}
\label{sec:rq2}

\subsubsection{Component Contributions}

\providecommand{\tool}{\textsc{Brief}}
\providecommand{\gptmodel}{GPT-6-Luna}

\begin{table*}[t]
\centering
\caption{Component and evidence-source ablations.}
\label{tab:ablation-results}
\small
\begin{adjustbox}{max width=\linewidth}
\begin{tabular}{lrrrrrrr}
\toprule
Configuration & \#Plausible & \#Correct & $\Delta$\#Correct & Cost (\$) & Avg. Input (K) & Avg. Output (K) & Repair Iterations \\
\midrule
\multicolumn{8}{l}{\emph{Component ablation}} \\
Full \tool & 336 & 256 & -- & 0.057 & 221.5 & 71.4 & 34.7 \\
\quad w/o Plan Assessment & 287 & 216 & $-$40 ($-$15.6\%) & 0.059 & 259.2 & 67.1 & 35.6 \\
\quad w/o Plan Synthesis & 191 & 127 & $-$129 ($-$50.4\%) & 0.098 & 546.0 & 86.5 & 36.3 \\
\quad Baseline & 74 & 59 & $-$197 ($-$77.0\%) & 0.143 & 986.1 & 88.7 & 39.7 \\
\midrule
\multicolumn{8}{l}{\emph{Evidence-source ablation}} \\
\quad w/o NL Documentation & 172 & 113 & $-$143 ($-$55.9\%) & 0.068 & 324.7 & 71.5 & 35.9 \\
\quad w/o Code Context & 120 & 91 & $-$165 ($-$64.5\%) & 0.061 & 360.1 & 50.9 & 38.7 \\
\quad w/o Failing Tests & 107 & 82 & $-$174 ($-$68.0\%) & 0.116 & 644.0 & 104.0 & 38.6 \\
\bottomrule
\end{tabular}
\end{adjustbox}
\end{table*}

Table~\ref{tab:ablation-results} reports component and evidence-source
ablations of \brief with RepairAgent under \gptmodel{}. For each
configuration, we report plausible and correct repairs, the change in correct
repairs relative to Full \brief ($\Delta$\#Correct), average per-bug cost,
input and output token usage, and repair iterations.

\begin{samepage}
The component ablation progressively removes stages from the full pipeline:
\begin{itemize}
    \item \emph{w/o Plan Assessment} supplies the synthesized Repair Plan
    directly to the backend, bypassing assessment and refinement.
    \item \emph{w/o Plan Synthesis} supplies the collected evidence directly
    to the backend, bypassing both plan synthesis and plan assessment.
    \item \emph{Baseline} uses the original backend without \brief's evidence
    collection, plan synthesis, or assessment.
\end{itemize}
\end{samepage}

The component variants are cumulative, so adjacent differences depend on
removal order and do not isolate independent effects. Each evidence-source
variant instead removes one source from Full \brief while keeping the
pipeline stages intact.

\textbf{Component ablation.}
Full \brief achieves 336 plausible and 256 correct repairs. Viewed in the
forward direction, the cumulative ablation reveals a clear progression:
the baseline yields 59 correct repairs, direct evidence injection raises this
to 127 (+68), synthesizing the same evidence into an unchecked Repair Plan
further increases it to 216 (+89, +70.1\%), and assessment and refinement
raise the count to 256 (+40, +18.5\%). This progression shows that additional
evidence is useful, but structuring that evidence into explicit repair intent
provides the largest adjacent gain, while assessment delivers a further
improvement over an already structured plan.

The same trend appears in resource use. Full \brief has the lowest average
per-bug cost (\$0.057), input-token usage (221.5K), and repair iterations
(34.7) among the four configurations. In contrast, raw evidence injection
costs \$0.098 per bug, consumes 546.0K input tokens, and requires 36.3 repair
iterations on average, while the baseline costs \$0.143, consumes 986.1K input
tokens, and requires 39.7 iterations. Bypassing assessment saves only 4.3K
output tokens on average (67.1K vs.\ 71.4K) while losing 40 correct repairs.

One plausible explanation is that the assessed Repair Plan gives the backend a
more focused repair intent, reducing both the amount of raw context it must
process and the number of repair attempts needed before termination. The lower
iteration counts are consistent with faster convergence of the repair process,
although our aggregate measurements do not isolate this mechanism or measure
wall-clock latency directly. Empirically, the assessed Repair Plan provides a
more effective and resource-efficient interface than raw evidence injection in
this configuration.

\textbf{Evidence-source ablation.}
Removing NL documentation, code context, or failing tests reduces correct
repairs from 256 to 113, 91, and 82, corresponding to losses of 55.9\%,
64.5\%, and 68.0\%, respectively. Thus, none of the three evidence sources is
redundant in the full pipeline.

Interestingly, all three ablations also increase average per-bug cost,
input-token usage, and repair iterations. The largest degradation occurs
without failing tests, where correct repairs drop by 174 and cost rises from
\$0.057 to \$0.116. These results suggest that informative evidence not only
improves repair effectiveness but may also reduce downstream exploration.

\subsubsection{Plan Assessment and Refinement}
\label{sec:assess_experiment}

\providecommand{\tool}{\textsc{Brief}}
\providecommand{\gptmodel}{GPT-6-Luna}

\begin{table*}[t]
\centering
\caption{Plan acceptance and repair outcomes. For rejected plans,
\brief provides no repair guidance and falls back to the
original backend; therefore the reported correct repairs correspond to the
baseline outcome.}
\label{tab:plan-accept}
\small
\setlength{\tabcolsep}{2pt}
\renewcommand{\arraystretch}{1.12}
\begin{tabular*}{\linewidth}{@{\extracolsep{\fill}}llrrrrrrr@{}}
\toprule
\multirow{3}{*}{Model} & \multirow{3}{*}{Bugs} & \multirow{3}{*}{\#Planned} & \multirow{3}{*}{\#Accepted}
 & \multicolumn{3}{c}{Accepted plans} & \multicolumn{2}{c}{Rejected plans} \\
\cmidrule(lr){5-7}\cmidrule(lr){8-9}
 & & & & Baseline & \multicolumn{2}{c}{+\tool{}} & \multirow{2}{*}{\#Bugs} & Fallback \\
\cmidrule(lr){6-7}
 & & & & \#Correct & \#Plausible & \#Correct & & \#Correct \\
\midrule
\multirow{3}{*}{DeepSeek-V3.2} & All & 835 & 246 (29.5\%) & 21 & 191 & \textbf{159} & 589 & 45 \\
 & Single-hunk & 463 & 172 (37.1\%) & 20 & 151 & \textbf{129} & 291 & 40 \\
 & Multi-hunk & 372 & 74 (19.9\%) & 1 & 40 & \textbf{30} & 298 & 5 \\
\midrule
\multirow{3}{*}{\gptmodel} & All & 835 & 813 (97.4\%) & 59 & 336 & \textbf{256} & 22 & 0 \\
 & Single-hunk & 463 & 456 (98.5\%) & 54 & 251 & \textbf{202} & 7 & 0 \\
 & Multi-hunk & 372 & 357 (96.0\%) & 5 & 85 & \textbf{54} & 15 & 0 \\
\bottomrule
\end{tabular*}
\end{table*}

Table~\ref{tab:plan-accept} separates plan acceptance from downstream repair
outcomes for RepairAgent. All 835 bugs enter planning under each model. Bugs
receiving an Accepted Repair Plan are compared with the baseline on the same
subset; when no plan is accepted within three rounds, \brief discards the
candidate guidance and falls back to the original backend.

Plan acceptance differs sharply across models. Under DeepSeek-V3.2, 246 of
835 plans are accepted (29.5\%), with a higher acceptance rate for single-hunk
than multi-hunk bugs (37.1\% vs.\ 19.9\%). Under \gptmodel{}, 813 plans are
accepted (97.4\%), and acceptance remains high for both single-hunk and
multi-hunk bugs (98.5\% and 96.0\%, respectively). This contrast shows that
the assessor's acceptance behavior is strongly model-dependent, particularly
for more complex multi-hunk repairs.

On bugs with accepted plans, +\brief substantially outperforms the original
backend. Under DeepSeek-V3.2, it produces 159 correct repairs versus 21 for
the baseline on the same 246 bugs (+138); under GPT-6-Luna, it produces 256
versus 59 (+197). The gains cover both complexity classes: 109 single-hunk
and 29 multi-hunk repairs under DeepSeek-V3.2, and 148 and 49, respectively,
under GPT-6-Luna.

Rejected plans do not terminate repair. For the 589 DeepSeek-V3.2 cases in
which no plan is accepted, fallback to the original backend preserves 45
correct repairs; together with the 159 repairs obtained on accepted-plan bugs,
this yields the 204 total correct repairs reported in
Table~\ref{tab:full-benchmark-results}. Under \gptmodel{}, none of the 22
rejected cases is correctly repaired by the baseline, so the final total
remains 256. This fallback highlights the role of assessment as a gate on
\brief's guidance rather than a gate on whether the backend may attempt repair.

Importantly, plan acceptance is distinct from patch correctness. Among bugs
with accepted plans, 159/246 (64.6\%) are correctly repaired under
DeepSeek-V3.2 and 256/813 (31.5\%) under \gptmodel{}. In particular, the high
acceptance rate under \gptmodel{} does not translate into an equally high
correct-repair rate. Acceptance therefore indicates that a plan satisfies
\brief's evidence-based assessment criteria, not that the downstream patch is
guaranteed to be correct.

\providecommand{\gptmodel}{GPT-6-Luna}

\definecolor{baseorange}{RGB}{244,164,96}
\definecolor{oursblue}{RGB}{70,130,180}
\pgfplotsset{
  evalchart/.style={
    ybar=0pt, clip=false,
    width=0.46\textwidth, height=5.4cm,
    ymin=0,
    ymajorgrids, grid style={dashed, gray!35},
    axis lines*=left, axis line style={gray!70},
    xtick style={draw=none},
    tick label style={font=\small}, title style={font=\small, yshift=-2pt},
    ylabel style={font=\small},
    unbounded coords=discard,
    nodes near coords, nodes near coords style={font=\scriptsize},
    legend style={draw=none, font=\small, legend columns=2,
                  /tikz/every even column/.append style={column sep=14pt}},
    area legend,
  },
  hunkchart/.style={evalchart, bar width=24pt, ymax=200, ytick distance=25,
    symbolic x coords={Single-hunk,Multi-hunk}, xmin=Single-hunk, xmax=Multi-hunk,
    xtick={Single-hunk,Multi-hunk}, enlarge x limits=0.5},
  basebar/.style={fill=baseorange, draw=none,
                  nodes near coords style={font=\scriptsize, text=gray!60!black}},
  oursbar/.style={fill=oursblue, draw=none,
                  nodes near coords style={font=\scriptsize\bfseries, text=oursblue}},
}
\tikzset{gainlabel/.style={font=\scriptsize\bfseries, text=oursblue,
                           anchor=south west, xshift=24pt, yshift=4pt}}

\begin{figure*}[t]
\centering
\ref{roundlegend}\\[4pt]
\begin{adjustbox}{max width=\linewidth}
\begin{tikzpicture}
\begin{groupplot}[group style={group size=3 by 1, horizontal sep=1.8cm}, evalchart,
  /pgf/bar width=20pt, ymax=700, ytick distance=100, symbolic x coords={Round 1,Round 2,Round 3}, xmin=Round 1, xmax=Round 3, xtick={Round 1,Round 2,Round 3}, enlarge x limits=0.25]
\nextgroupplot[title={DeepSeek-V3.2}, ylabel={\# Bugs}, legend to name=roundlegend]
\addplot[basebar] coordinates {(Round 1,120) (Round 2,76) (Round 3,50)};
\addplot[oursbar] coordinates {(Round 1,92) (Round 2,45) (Round 3,22)};
\legend{Accepted Plans, Correct Repairs}
\nextgroupplot[title={\gptmodel}]
\addplot[basebar] coordinates {(Round 1,597) (Round 2,174) (Round 3,42)};
\addplot[oursbar] coordinates {(Round 1,214) (Round 2,36) (Round 3,6)};
\nextgroupplot[title={Sampled Bugs}, ylabel={\% of Sampled Bugs},
  /pgf/bar width=20pt, ymin=0, ymax=20, ytick distance=5,
  symbolic x coords={R1,R2,R3,R4,R5,R6}, xmin=R1, xmax=R6, xtick={R1,R2,R3,R4,R5,R6}, enlarge x limits=0.12,
  nodes near coords={\pgfmathprintnumber[fixed, fixed zerofill, precision=1]{\pgfplotspointmeta}\%}]
\addplot[basebar] coordinates {(R1,16.7) (R2,6.7) (R3,10.0) (R4,1.7) (R5,1.7) (R6,0.0)};
\end{groupplot}
\end{tikzpicture}
\end{adjustbox}
\caption{Plan acceptance and correct repairs across assessment rounds; the right panel shows the pilot for refinement-budget selection.}
\label{fig:rounds}
\end{figure*}
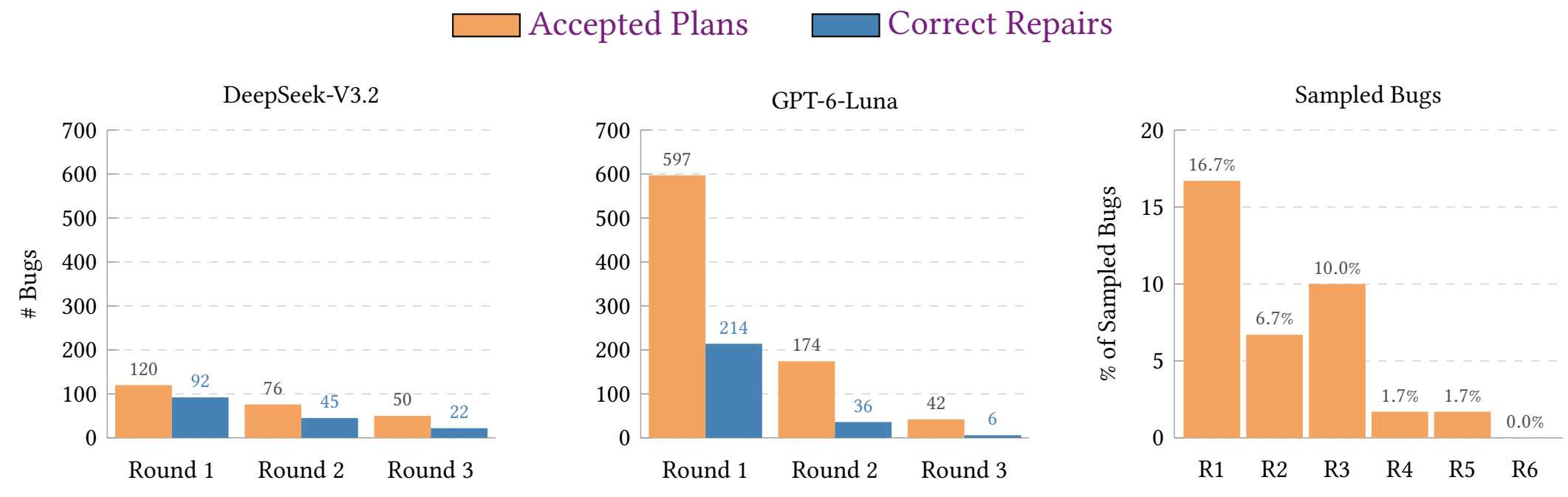

\textbf{Assessment and refinement across rounds.}
Figure~\ref{fig:rounds} examines when plans become acceptable during
refinement. Under DeepSeek-V3.2, 120 plans are accepted in Round~1, while
Rounds~2 and~3 recover another 76 and 50 plans; these later-round plans yield
45 and 22 correct repairs, respectively. Thus, 126 of the 246 accepted plans
(51.2\%) and 67 of the 159 correct repairs on accepted-plan bugs (42.1\%) come
from plans that are accepted only after refinement. Under GPT-6-Luna, 597 plans
are accepted initially, followed by 174 and 42 in Rounds~2 and~3, yielding
another 36 and 6 correct repairs.
These results show that refinement yields additional accepted plans, and that
plans accepted in later rounds still lead to correct repairs.

The pilot analysis on the right provides the basis for the three-round
refinement budget. Round~3 still yields accepted plans for 10.0\% of the
sampled bugs, while the additional contribution drops to 1.7\%, 1.7\%, and
0\% in Rounds~4--6. Based on this diminishing-return pattern, we use three
rounds in the main evaluation.

\begin{finding}
Transforming raw evidence into a Repair Plan increases correct repairs from
127 to 216 (+70.1\%), and assessment and refinement further increase them to
256 (+18.5\%). Removing any evidence source substantially degrades repair
effectiveness. Moreover, 67 and 42 correct repairs under DeepSeek-V3.2 and
GPT-6-Luna, respectively, come from plans accepted only after refinement.
\end{finding}

\subsection{RQ3: Repair Cost and Efficiency}
\label{sec:rq3}

\providecommand{\tool}{\textsc{Brief}}
\providecommand{\gptmodel}{GPT-6-Luna}
\providecommand{\backendB}{ReinFix}

\begin{table*}[t]
\centering
\caption{Overall repair cost and resource use.}
\label{tab:cost-token}
\small
\begin{adjustbox}{max width=\linewidth}
\begin{tabular}{lllrrrr}
\toprule
Backend & Model & Configuration & Cost (\$) & Avg. Input (K) & Avg. Output (K) & Repair Iterations \\
\midrule
\multirow{4}{*}{RepairAgent} & \multirow{2}{*}{DeepSeek-V3.2} & Baseline & 0.075 & 256.4 & 7.9 & 35 \\
 & & +\tool & 0.074 & 249.9 & 10.8 & 31.7 \\
\cmidrule{2-7}
 & \multirow{2}{*}{\gptmodel} & Baseline & 0.143 & 986.1 & 88.7 & 39.7 \\
 & & +\tool & 0.057 & 221.5 & 71.4 & 34.7 \\
\midrule
\multirow{4}{*}{\backendB} & \multirow{2}{*}{DeepSeek-V3.2} & Baseline & 0.042 & 129.6 & 14.6 & 8.1 \\
 & & +\tool & 0.045 & 137.9 & 15.7 & 7.7 \\
\cmidrule{2-7}
 & \multirow{2}{*}{\gptmodel} & Baseline & 0.080 & 427.2 & 75.2 & 8.4 \\
 & & +\tool & 0.042 & 154.4 & 53.2 & 7.3 \\
\bottomrule
\end{tabular}
\end{adjustbox}
\end{table*}

Table~\ref{tab:cost-token} compares the end-to-end resource consumption of
the baseline and +\brief configurations over all 835 bugs. We report average
per-bug cost, input and output token usage, and repair iterations to characterize
the computational efficiency of each configuration.

\textbf{Overall resource use.}
Average per-bug cost decreases in three of the four backend--model
configurations. Under \gptmodel{}, cost drops from \$0.143 to \$0.057
with RepairAgent (-60.1\%) and from \$0.080 to \$0.042 with ReinFix
(-47.5\%). Under DeepSeek-V3.2, RepairAgent remains nearly cost-neutral
(\$0.075 vs.\ \$0.074), while ReinFix increases modestly from \$0.042
to \$0.045 (+7.1\%). Since correct repairs increase in all four
configurations (Table~\ref{tab:full-benchmark-results}), only ReinFix with
DeepSeek-V3.2 exhibits an effectiveness--cost tradeoff.

The same pattern is reflected in token usage. Under \gptmodel{}, +\brief
reduces input tokens from 986.1K to 221.5K with RepairAgent and from
427.2K to 154.4K with ReinFix; output tokens decrease from 88.7K to
71.4K and from 75.2K to 53.2K, respectively. Under DeepSeek-V3.2, token
changes are smaller and mixed. Repair iterations nevertheless decrease
in all four configurations: from 35.0 to 31.7 and 39.7 to 34.7 with
RepairAgent, and from 8.1 to 7.7 and 8.4 to 7.3 with ReinFix.

Overall, these results show that \brief improves repair effectiveness
without a proportional resource penalty and, in three of the four
configurations, also reduces average per-bug cost.

\providecommand{\tool}{\textsc{Brief}}
\providecommand{\gptmodel}{GPT-6-Luna}
\providecommand{\backendB}{ReinFix}

\begin{table}[htbp]
\centering
\caption{Average per-bug repair cost and resource use by plan outcome.
Rejected-plan values exclude baseline fallback; iterations are reported
only for accepted plans.}
\label{tab:cost-plan-outcome}
\small
\setlength{\tabcolsep}{3pt}
\begin{tabular*}{\linewidth}{@{\extracolsep{\fill}}lllrrrrr@{}}
\toprule
Outcome & Backend & Model & \#Bugs & Cost (\$) & Input (K) & Output (K) & Iterations \\
\midrule
\multirow{4}{*}{\emph{Accepted}} & \multirow{2}{*}{RepairAgent} & DeepSeek-V3.2 & 246 & 0.056 & 185.6 & 10.1 & 23.9 \\
 & & \gptmodel & 813 & 0.055 & 200.4 & 70.7 & 34.6 \\
\addlinespace[2pt]
 & \multirow{2}{*}{\backendB} & DeepSeek-V3.2 & 245 & 0.035 & 109.8 & 10.5 & 6.8 \\
 & & \gptmodel & 812 & 0.041 & 146.2 & 52.4 & 7.3 \\
\midrule
\multirow{4}{*}{\emph{Rejected}} & \multirow{2}{*}{RepairAgent} & DeepSeek-V3.2 & 589 & 0.007 & 20.3 & 3.2 & -- \\
 & & \gptmodel & 22 & 0.005 & 15.1 & 7.5 & -- \\
\addlinespace[2pt]
 & \multirow{2}{*}{\backendB} & DeepSeek-V3.2 & 590 & 0.007 & 20.0 & 3.2 & -- \\
 & & \gptmodel & 23 & 0.006 & 16.3 & 7.9 & -- \\
\bottomrule
\end{tabular*}
\end{table}

Table~\ref{tab:cost-plan-outcome} decomposes +\brief's resource use by plan
outcome. For accepted plans, the reported values include downstream repair
with the Accepted Repair Plan. For rejected plans, they include only \brief's
processing before fallback; the subsequent execution of the original backend
is included in the end-to-end averages in Table~\ref{tab:cost-token}.

The key difference across models is the frequency of fallback. Under
DeepSeek-V3.2, 589 of 835 RepairAgent plans and 590 of 835 ReinFix plans are
rejected. Each rejected case incurs about \$0.007 of \brief processing before
the original backend is still invoked, yielding roughly \$0.005 of
pre-fallback overhead per benchmark bug when averaged over all 835 bugs.
This frequent fallback offsets part of the resource savings achieved when an
Accepted Repair Plan is available.

Under \gptmodel{}, by contrast, only 22 RepairAgent plans and 23 ReinFix plans
are rejected, so the cumulative fallback overhead is negligible. This pattern
helps explain the end-to-end results in Table~\ref{tab:cost-token}: +\brief
achieves large cost reductions under \gptmodel{}, whereas the savings are
attenuated under DeepSeek-V3.2. Thus, \brief's overall efficiency depends not
only on the cost of producing repair guidance, but also on how often assessment
can deliver an accepted plan and avoid fallback.

\begin{finding}
\brief improves repair effectiveness without proportional resource overhead:
average per-bug cost decreases in three of four backend--model configurations,
and repair iterations decrease in all four. The largest cost reductions occur
under GPT-6-Luna (60.1\% with RepairAgent and 47.5\% with ReinFix), where
fallback is rare. Under DeepSeek-V3.2, frequent fallback adds pre-fallback
overhead and attenuates these savings.
\end{finding}

\section{Discussion}
\label{sec:discussion}

\noindent\textbf{Repair intent as an intermediate software artifact.}
Our results suggest that repair intent need not remain implicit inside a
backend's reasoning trajectory. Externalizing it as a structured artifact
creates a boundary between understanding a defect and realizing its fix: the
plan captures what behavioral obligations the repair should satisfy, while the
backend remains responsible for how those obligations are implemented. This
separation makes repair intent inspectable, assessable, and reusable across
different repair procedures, suggesting a broader design space in which repair
reasoning and patch generation can evolve independently.

\noindent\textbf{Assessment should be selective rather than mandatory guidance.}
A synthesized plan is itself a potential source of error. \brief therefore
treats assessment as a control layer that decides whether synthesized guidance
should influence repair at all. Rejected plans are withheld and the backend
falls back to its original procedure. This design suggests that future
LLM-based repair systems may benefit not only from generating better guidance,
but also from learning when that guidance is sufficiently grounded to be used.
The strong variation in acceptance behavior across models further indicates
that such control mechanisms may need to be calibrated to the underlying LLM.

\noindent\textbf{The interface is only as reliable as its evidence and assessor.}
Externalizing repair intent does not eliminate uncertainty. Incomplete or
misleading evidence can produce an incomplete plan, and an assessor using the
same underlying model may share the synthesizer's blind spots. Moreover,
frequent rejection preserves the original repair path but introduces additional
pre-fallback cost. These observations motivate stronger evidence-aware
verification, independently trained or heterogeneous assessors, and adaptive
refinement budgets as directions for future work.

\section{Threats to Validity}
\label{sec:threats}

\noindent\textbf{Internal validity.}
LLM-based repair is nondeterministic, so repeated executions may produce
different plans, assessment decisions, and patches. Our paired comparisons keep
the backend, model, repair inputs, execution environment, and repair budget fixed
between the baseline and \brief configurations, reducing confounding factors but
not eliminating run-to-run variance. Results across DeepSeek-V3.2 and
GPT-6-Luna provide evidence across two models, while repeated trials would be
needed to quantify stochastic variability more precisely. Patch correctness
also involves human judgment: plausible patches that do not syntactically match
the developer patch are manually inspected against the intended behavior.

\noindent\textbf{Construct validity.}
Acceptance
by the Plan Assessor does not establish that a Repair Plan or downstream patch is
correct. The synthesizer and assessor use the same underlying LLM in separate
invocations and may therefore share model biases or misinterpret the same
evidence. Grounding, target consistency, coverage, and concreteness operationalize
specific properties of repair guidance rather than formal correctness. 

\noindent\textbf{External validity.}
We evaluate \brief on 835 Java bugs from 17 Defects4J projects using two repair
backends and two LLMs. The results therefore do not establish generality to
other languages, ecosystems, models, or repair systems, although the
cross-backend evaluation provides evidence that the same repair-intent interface
can augment distinct repair procedures. Defects4J is a public benchmark, and its
programs, tests, or patches may appear in model training data. Paired comparisons
expose baseline and \brief configurations to the same benchmark, but cannot
eliminate contamination risk. We additionally evaluate both backends with DeepSeek-V3.2
on 41 SWE-bench-Live/MultiLang Java bugs created after that model's release.
The gains on these bugs (Section~\ref{sec:post-cutoff}) reduce concern
that \brief's Defects4J results rely on memorization of specific fixes,
although this evaluation cannot rule out other leakage channels or establish
generality beyond the evaluated setting. Finally, \brief depends on the quality of available
evidence; sparse, ambiguous, or outdated documentation, tests, or code context
may limit the quality of synthesized repair intent.

\section{Related Work}
\label{sec:related-work}

\noindent\textbf{Diagnosis and structured reasoning for repair.}
LLMs can plan before generation, interleave reasoning with actions, and refine
their outputs through self-feedback or critique
~\cite{wei2022chain,wang2023plan,yao2022react,madaan2023self,
shinn2023reflexion,gou2024critic}. Similar ideas have been applied to code
generation and debugging
~\cite{jiang2024self,li2025structured,chen2024teaching,olausson2024self}.
Recent LLM-based APR systems make such pre-repair reasoning explicit.
PracRepair combines static and dynamic evidence with question-driven diagnosis
and summarizes the analysis into repair hypotheses
~\cite{cheng2026pracrepair}; CAMP structures pre-repair reasoning around
domain knowledge, intended functionality, and bug causes~\cite{wu2026camp};
DRCodePilot derives a solution design before patching
~\cite{zhao2024enhancing}; and SpecRover extracts intended behavior to guide
patch generation and review~\cite{ruan2025specrover}. Other work explores
self-directed repair reasoning~\cite{yin2024thinkrepair} and semi-formal
reasoning over explicit premises and conclusions~\cite{ugare2026agentic}.

\brief differs in the role assigned to pre-repair reasoning. Rather than
treating it as internal state of an end-to-end repair workflow, \brief
externalizes repair intent as a backend-decoupled interface. The resulting
Repair Plan is checked against its source evidence for grounding, target
consistency, coverage, and concreteness before it can influence patch
generation, and only accepted guidance is supplied to an existing repair
backend. Thus, \brief focuses not only on producing repair guidance, but also
on assessing that guidance before downstream use.

\noindent\textbf{Automated program repair.}
Traditional APR includes search- and template-based methods
~\cite{legoues2012genprog,liu2019tbar,kim2013automatic},
constraint-based synthesis
~\cite{nguyen2013semfix,xuan2016nopol,mechtaev2016angelix},
and learning-based repair
~\cite{lutellier2020coconut,chen2021sequencer,li2020dlfix}.
LLM-based approaches generate patches from buggy code and context
~\cite{xia2023automated,jiang2023impact}, incorporate execution feedback
~\cite{kang2025explainable,ye2024iter}, use conversational repair
~\cite{xia2024automated}, or provide semantic and multi-granularity guidance
~\cite{lin2024one,pabba2025semagent}. Agentic repair systems further navigate
repositories, edit code, and execute tests through tool interfaces
~\cite{yang2024swe,zhang2024autocoderover,wang2025openhands}.
RepairAgent~\cite{bouzenia2025repairagent} and
ReinFix~\cite{zhang2025reinfix}, which we use as downstream backends, represent
distinct repair procedures.

\brief is complementary to these patch-generation systems. It does not replace
their search, editing, validation, or refinement procedures; instead, it
provides an assessed statement of repair intent as structured context before
patch generation. This separation allows the same repair-intent interface to
augment different backends without redesigning their native repair loops.

\noindent\textbf{Behavioral and specification guidance.}
Behavioral information has long been used to constrain repair. Formal
preconditions, postconditions, and invariants can guide localization and repair
~\cite{al2024invariant}; temporal properties can serve as behavioral oracles
~\cite{song2024provenfix}; and specification-guided repair constrains candidate
patches using formal behavioral requirements
~\cite{wu2025specification,pei2014autofix}. Recent work also derives
intermediate behavioral signals from execution feedback
~\cite{le2026enhancing} or performs behavior-centric repair reasoning
~\cite{zhu2026specification}.

\brief likewise exploits behavioral information, but does not require a formal
specification. It synthesizes repair intent from complementary project evidence,
including documentation, failing tests, and code context, and then checks the
resulting guidance against those same sources before patch generation. The
Repair Plan therefore serves as evidence-backed guidance rather than a formal
correctness specification.

\noindent\textbf{Program evidence for repair.}
Fault localization identifies program elements associated with observed
failures~\cite{wong2016survey}, including through LLM-based analysis of code
and failure reports~\cite{kang2024autofl}. Localization-oriented pipelines
make this stage explicit: Agentless localizes hierarchically from files to edit
locations~\cite{xia2024agentless}, RGFL ranks files and code elements using
bug-specific explanations~\cite{sepidband2026rgfl}, SynFix retrieves dependency
context from a code relation graph~\cite{tang2025synfix}, and SHERLOC supplies
repair agents with locations and diagnostic findings
~\cite{tamoyan2026sherloc}. Developer documentation provides another source of
behavioral evidence and has been used for inconsistency detection
~\cite{tan2007icomment,tan2012tcomment}, specification synthesis
~\cite{zhai2020c2s,endres2024nl2postcond,xie2023impact}, and test generation
~\cite{goffi2016toradocu,blasi2018jdoctor}.

These approaches demonstrate the value of identifying and retrieving relevant
repair evidence. \brief addresses a different stage: given supplied repair
locations and heterogeneous project evidence, it synthesizes their implications
into explicit repair intent and assesses that intent before downstream repair.
Accordingly, \texttt{fix\_targets} connect supplied locations to behavioral
requirements; they do not constitute an independent fault-localization
technique.

\section{Conclusion}
\label{sec:conclusion}

We introduced \brief, an evidence-guided repair harness that externalizes
repair intent as a backend-decoupled, evidence-assessed interface between
project evidence and patch generation. \brief synthesizes documentation,
failing tests, and code context into a structured Repair Plan, assesses and
refines the plan before use, and supplies only accepted guidance to existing
repair backends while preserving their native repair procedures.

Across 835 Defects4J bugs, two repair backends, and two LLMs, \brief improves
correct repairs in all four backend--model configurations by 117--197 bugs,
corresponding to relative gains of 60.9\%--333.9\%. Ablation results further show a clear progression: direct evidence injection
yields 127 correct repairs, structuring that evidence into a Repair Plan raises
the count to 216 (+70.1\%), and assessment and refinement further increase it
to 256 (+18.5\%).
These gains do not require proportional resource overhead, with average
per-bug cost decreasing in three of four configurations and repair iterations
decreasing in all four. Together, these results provide evidence that assessed
repair intent can serve as an effective interface for augmenting distinct
LLM-based repair procedures.

\section{Data Availability}
The anonymous replication package contains the source code and experimental results needed to inspect and reproduce this study. It is publicly available for review at \url{https://anonymous.4open.science/r/spec-repair-test-38EA/}.

\bibliographystyle{ACM-Reference-Format}
\bibliography{custom}

\end{document}